\documentclass[12pt]{article}
\usepackage{amsmath,amssymb,amsfonts,epsfig,graphicx}

\newcommand{\bse}{\begin{subequations}}
\newcommand{\ese}{\end{subequations}}
\newcommand{\be}{\begin{equation}}
\newcommand{\ee}{\end{equation}}
\newcommand{\bea}{\begin{eqnarray}}
\newcommand{\eea}{\end{eqnarray}}
\newcommand{\ba}{\begin{array}}
\newcommand{\ea}{\end{array}}

\makeatletter \@addtoreset{equation}{section}

\makeatother

\def\A0{{\cal A}_0}

\def\L5{{\cal L}_5}

\begin{document}

%\rightline{} \vspace{1cm}
%\begin{center}

\baselineskip 18pt%

\begin{titlepage}
\vspace*{1mm}%
\hfill%
\vbox{
    \halign{#\hfil        \cr
           hep-th/0612181 \cr
           IPM/P-2006/065 \cr
           IC/2006/132\cr
    } % end of \halign
      }  % end of \vbox
\vspace*{10mm}%
\begin{center}
\centerline{{\Large {\bf Non-Abelian Magnetized Blackholes and Unstable Attractors
%Instabilities in Magnetized  Einstein-Yang-Mills Solutions
}}} %\vspace*{-3cm}

%\centerline{{\large{\bf Instability of EYM solutions in Background Magnetic Flux}}}
%\centerline{{\large{\bf and  its Relevance to String Theory}}}

\vspace*{0.6cm}
{\bf\large A.E. Mosaffa$^1$, S.~Randjbar-Daemi$^2$, M. M. Sheikh-Jabbari$^1$}%
\vspace*{0.4cm}

$^1${\it {Institute for Studies in Theoretical Physics and Mathematics (IPM)\\
P.O.Box 19395-5531, Tehran, IRAN}}\\
{E-mails: {\tt mosaffa, jabbari@theory.ipm.ac.ir}}\\
$^2${\it{ The Abdus Salam International Centre for Theoretical Physics,\\ Strada Costiera 11 34014, Trieste, Italy}}\\
{E-mail: {\tt seif@ictp.trieste.it}}
\vspace*{1.5cm}
\end{center}

\begin{center}{\bf Abstract}\end{center}
\begin{quote}
Fluctuations of non-Abelian gauge fields in a background
magnetic flux contain tachyonic modes and hence the background is
unstable. We extend these results to the cases where the background
flux is coupled to Einstein gravity and show that the corresponding
spherically symmetric geometries,
 which in the absence of a cosmological constant
are of the form of Reissner-Nordstr\"om blackholes or the
$AdS_2\times S^2$, are also unstable. We discuss the relevance of
these instabilities to several places in string theory including
various string compactifications and the attractor mechanism. Our
results for the latter imply that the attractor mechanism shown to
work for the extremal Abelian charged blackholes, cannot be applied
  in a straightforward way to the
extremal non-Abelian colored blackholes.

\end{quote}%

\end{titlepage}

\newpage

%\renewcommand{\theequation}{\thesection.\arabic{equation}}

%\tableofcontents

\section{Introduction and Motivation}

 Stability analysis of systems of charged particles coupled to Einstein gravity in
four dimensions is an old and well studied subject.  Static,
spherically symmetric solution with electric and/or magnetic charge,
the Reissner-Nordstr\"om blackhole, has been shown to be stable
under classical stability test. Charged or neutral scalars can also
be added to the Einstein-Maxwell theory. A special case of the
latter is the Einstein-Maxwell-Dilaton (EMD) theory. String theory
toroidal compactifications at  generic points in their moduli space
indeed lead to a EMD theory with several $U(1)$'s as well as
dilatonic fields (e.g. see \cite{Sen-94}). All these generalized
charged blackholes are stable.

 As the next generalization,  one may couple non-Abelian gauge theories
to Einstein gravity. These theories are hence called
Einstein-Yang-Mills (EYM) theories \cite{Bartnik}. Such theories
have also been studied in some detail and it has been shown that
they have an infinite discrete family of  globally regular solutions
as well as blackhole solutions  for any given value of the horizon
radius \cite{Perry75-7, SWY, Lavrelash}. Despite having time independent
colored globally regular or colored blackholes solutions, it has
been argued that all of these solutions are unstable. (For a more detailed review see
\cite{Volkov-Review}.)

When Dilaton or Higgs scalars are added to the system we obtain,
respectively,  EYMD or EYMH theories. These are the kind of theories
which can arise from string theory compactifications, though not in
generic points of the moduli space of toroidal compactifications.
They also arise from compactification on group manifolds or generic
Calabi-Yau manifolds. As we are interested in gravity theories
arising from string compactifications, in this note we shall focus
on these theories and review some of the results regarding the
stability of such charged blackholes in section \ref{section2.2}. We
extend the results on the instability of the colored blackholes to a
wider class in which the unstable (tachyonic) mode is not an \emph{s}-wave.

In the rest of this note, we will  first review the results of
\cite{RSS2,DRT}, stating the existence of the tachyonic modes in the
fluctuations of the Yang-Mills fields in a monopole background. In
section \ref{section2.1}, we complete the existing literature by
showing that indeed the existence of these tachyonic modes leads to
the instability of the background. In section \ref{section2.2}, we
review arguments in the GR literature in which it has been shown
that a similar class of systems, namely the colored blackhole
systems, are unstable. In section 3,  we focus on the instabilities
in the presence of gravitational field. In this section we show that
if a Yang-Mills configuration has destabilizing modes among its
fluctuations already in the flat space, then coupling to gravity
will not change the situation as regards to stability. To this end
first we show that the Dirac monopole in a EYM(D) theory generically
has fluctuations which are localized in space but grow exponentially
in time. With a given  monopole configuration or the corresponding
magnetic flux, the space time geometry is either of the form of
(colored) Reissner-Nordstr\"om (RN) blackhole or $AdS_2\times S^2$.
We show that  both of the two cases are suffering from a similar
kind of instability present in the non-gravitational case. The
solutions with minimal monopole charge are stable. This seems to
have been overlooked in previous analysis.

For the blackhole solutions, our results extend the known
instability arguments of the GR literature to a larger class, and in
particular to the cases different from \emph{s}-wave perturbations
and monopole charges different from minimal value for the tensorial
perturbations. The $AdS_2\times S^2$ is of relevance to the
attractor mechanism which is at work for extremal blackhole
solutions in the Einstein-Maxwell-Dilaton (EMD) theories (e.g. see
\cite{Attractor, Attractor-Triv}). The instability of the (extremal)
colored blackholes and its near horizon geometry, the $AdS_2\times
S^2$ background, implies that, if applicable at all, the attractor
mechanism in its present form does not work for the colored
non-Abelian blackholes. In section 4, we discuss that  the EYMD or
EYMH theories are very relevant to string theory and therefore the
results of the colored blackhole instabilities apply to string
theory compactifications. We end with concluding remarks and some
interesting open problems.

\section{Review of instabilities in background non-Abelian  magnetic flux}
This section is mainly a review of known results. In sections
\ref{section2.1} and \ref{section2.2}, respectively, we consider the
non-gravitational and gravitational cases and point out that in both
cases the fluctuations of non-Abelian gauge fields about a monopole
background contain unstable modes.

\subsection{Linearized analysis of the gauge fields fluctuations}\label{section2.1}

Fluctuations of a non-Abelian Yang-Mills field around a magnetic
monopole background on $R^{3+1}$ which has a non-vanishing flux through $S^2$, has tachyonic modes which can create
instabilities \cite{RSS2}.  To study such fluctuations it is
sufficient to study the linearized Yang-Mills equations $D_M F^{MN}
=0$, where $F^{MN}$ is the gauge field strength, around the solution
of interest. To proceed we use the notation  $M={\mu, m}$, where
$\mu$ ranges over $x^0=t$ and $x^1=r$ and $m$ ranges over $\theta$
and $\phi$. We also
take the four dimensional metric of the form%
\be \label{flatmet} ds^2 = g_{\mu\nu}dx^\mu dx^\nu + r^2
g_{mn}dy^mdy^n \ee where $g_{mn}$ denotes the standard metric on a
$S^2$ of unit radius.
%Here the indices are raised and lowered by the background metric.
Next we write%
\be%
W_M= \bar A_M +V_M %
\ee%
 where $W_M$ is the gauge field potential and $\bar A_M$ is the background
potential corresponding to a Dirac monopole located at the origin
(and hence has spherical symmetry)
\be%
\label{monopole}
\bar A =\frac{1}{2} n^iT_i(cos\theta -1)d\phi%
\ee%
where $T_i$ is a basis of matrices in the  Cartan sub-algebra of
the Lie algebra of $G$ and, in an appropriate normalization, $n_i$
which correspond to the charge(s) of the monopole are integers.
Yang-Mills equations, up to linear order in $V_n$,  give,%
\be \nabla_\mu^2 V^n + \frac{1}{r^2} D_m^2 V^n - [D_m, D_n] V^m -
\frac{i}{r^2}[V_m, \bar
F^{mn}] =0 .%
\ee%
In the above the covariant derivatives $\nabla^\mu$ and $\nabla^m$
are  respectively constructed using the metrics $g_{\mu\nu}$ and
$g_{mn}$ and the $m,n$ and $\mu,\nu$ indices are raised and lowered
by  metrics $g_{mn}$ and $g_{\mu\nu}$.
 $D_m$ is defined by%
\be%
D_m V_n =  \nabla_m V_n -i[\bar A_m, V_n],%
\ee%
and
\be\label{BG-F}
\bar F=-\frac{n_i}{2}T^i\sin\theta d\theta\wedge d\phi. \ee
The above equations of motion are written in the Lorentz gauge%
\be%
\nabla_{\mu} V^\mu + \frac{1}{r^2}D_{m}V^m =0.%
\ee%
Our interest is primarily in the  equation for $V^n$.  We can
express the commutators in terms of the curvature tensor. The
equation simplifies to%
\be%
\nabla^2 V^n + \frac{1}{r^2}\{ ( D_m^2-1) V^n  -2 i[V_m, \bar
F^{mn}] \}=0.%
\ee%

 Expanding  $V^n( t, r,\theta,\phi)$ in harmonics on $S^2$ yields
an infinite number of fields in the $1+1$ dimensional space spanned
by $t$ and $r$.  The general formalism has been given in
\cite{RSS1}. To perform such an expansion  it is necessary to use
complex basis in the tangent space of $S^2$ and denote the component
of $V_n$  with respect to such basis by $V_+ $ and $V_-$. The
equations then  separate and we
obtain%
\be \nabla^2 V_+ + \frac{1}{r^2}\{ ( D_m^2 -1) V_+  -2 i[V_+, \bar
F_{-+}] \}=0. \ee%
Writing
 \be
 V_+ = V_+^iT_i + V_+^aT_a,
 \ee
  where $T_i, T_a$ are, respectively, the generators in and outside the
  Cartan subalgebra
  satisfying %
  \be%
   [T_i, T_a] = \alpha_{ia} T_a,
   \ee%
   the equation for $V_{+a}$ becomes
   \be%
   \nabla^2 V_{+a} + \frac{1}{r^2}\{ ( D_m^2 -1) V_{+a}  +n^i\alpha_{ia}V_{+a}
   \}=0.
  \ee%
  This equation can be written in the more suggestive form of%
   \be\label{e.o.m-No-grav.}%
   \nabla^2 V_{+a} - \frac{1}{r^2} M^2V_{+a} =0
  \ee%
  where the ``mass operator'' $M^2$ is defined by%
  \be\label{mass-opt}%
  M^2 = - \{( D_m^2 -1)   + n^i\alpha_{ia}\}.
  \ee%

The spectrum of this operator is known. We know that for almost all
non-Abelian $G$'s it has a single negative eigenvalue (generically
corresponding to degenerate eigenstates)  and an infinite number of
positive eigenvalues \cite{RSS2, DRT}.  The multiplicity of the mode
with negative $M^2$, {\it tachyonic modes} \footnote{Note that if
the 2-dimensional metric is flat, i.e. $g_{\mu\nu}= \eta_{\mu\nu}$,
and the 4-dimensional space-time is a product of $R^2\times S^2$
then $M^2$ will be precisely the $mass^2$ operator in $R^2$ and the
appellation of tachyonic mode will be exact. By slight misuse of
language we shall continue to call $M^2$ the mass operator even for
a non flat $g_{\mu\nu}$ and for non factorized geometries. }, is
given by the dimension of the representation  of the overall
unbroken group to which they belong. For example, assume that the
gauge group is $SU(2)$. The background gauge field is that of a
monopole of charge $n$ which we take it to be a positive integer
greater than one in the direction of $T_3$.  The reason we exclude
$n=1$ is that there are no negative $M^2$ modes for this value of
$n$. In this case all the tangential fluctuations of the gauge field
are spinorial and the Abelian embedding is stable under small
fluctuations.

In this example, if there are no other symmetry breaking elements
(like charged scalar fields or the Higgs fields) the unbroken part
of the symmetry group in the $t,r$ space will be $U(1)\times
SU(2)_{KK}$, where, $U(1)$ is the unbroken subgroup of the gauge
group $SU(2)$ and the factor $SU(2)_{KK}$, is due to the spherical
symmetry of the background solution.  The tachyons in that case are
charged under the $U(1)$ and belong to the spin $j= n/2-1$
representation of $SU(2)_{KK}$. This in particular means that for
even $n$ the tachyons are in tensorial representations of the
rotation $SU(2)_{KK}$. Note that for odd $n$ the tachyons are
spinorial objects with respect to $SU(2)_{KK}$. The \emph{s}-wave
tachyon corresponds to $n=2$. Moreover there are  of $2j+1=n-1$
complex tachyons  (for $n>1$) with $ M^2= -n/2$ (for the tachyonic
modes $D^2_m=-j$ \cite{RSS2}).

One should, however, note that having negative $M^2$ is not in
itself sufficient to conclude the instability of the field
configuration, one needs also to analyze the propagation of these
negative $M^2$ modes in the $t, r$ space.

If we assume $g_{\mu\nu}=\eta_{\mu\nu}$, the fluctuation equation
\eqref{e.o.m-No-grav.} reduces to the simple wave equation
\be\label{e.o.m.opened} %
  -\partial^2_t V_{+}+ \partial^2_r V_{+} - \frac{1}{r^2} M^2V_{+}
  =0.
  \ee%
The most general solution to this equation turns out to be
\be\label{explicit.solution} %
V_{+}=e^{iEt} e^{iEr} r^{\beta}( C_1F(\beta, 2\beta; -2iEr) + C_2
(-2iEr)^{1-2\beta}F(1-\beta, 2-2\beta;-2iEr)
\ee%
where $F(a, c; x)$ is a confluent hypergeometric function and $C_1$
and $C_2$ are the two integration constants. $\beta$ is a solution
of
\be\label{beta-M} %
(\beta-1/2)^2=M^2+1/4.
\ee%
For $M^2<-1/4$, which is the case of interest for us, the right hand
side of this equation can be negative yielding a complex solution
for $\beta$. One can show that in this case there is a choice of the
integration constants which allow a normalizable solution with a
pure imaginary $E$. Such a solution will be localized in space but
delocalized (with exponential growth) in time and will signal an
instability.

If there are other charged fields in the model they may contribute
to the tachyonic mass and lift it to non-negative values. One such
possibility is adding the Higgs field in the adjoint (the Georgi-Glashow model) or the fundamental representations.
 For the former, there are
 {\it singular}  or {\it regular} ('t Hooft-Polyakov)  monopole
solutions.

Let us consider the Georgi-Glashow model, with the group $SU(2)$ and
a Higgs potential whose value vanishes in its minimum given by
$|\phi_{vac}|^2=v^2$ (e.g. see section 1.4 of
\cite{Luis-Hassan}):%
\be\label{GG-action}%
{\cal L}=-\frac{1}{4} \sum_{i=1}^3 \left(F^i_{MN}F^{i\
MN}+\frac{1}{2} D_M\phi^iD^M\phi^i\right)
-\frac{\lambda}{4}\left(\phi^2-v^2\right)^2 \ee%
where%
\be\label{G-Dphi-def}\begin{split}
F^i_{MN}&=\partial_{[M}W^i_{N]}-g_{YM} \epsilon^{ijk} W^j_MW^k_N\\%
D_M\phi^i&=\partial_M\phi^i-g_{YM}\epsilon^{ijk} W_M^j\phi^k.
\end{split}
\ee%
Our arguments in an obvious way generalizes to a generic
non-Abelian gauge group $G$. The {\it singular} monopole solution
is given by (see eq.(23) of \cite{Luis-Hassan})%
\be\label{singular-monopole-Higgs}
\begin{split}
\vec{\phi}_{vac}\cdot \vec{\phi}_{vac}&=v^2, \\
\vec{W}_M=\frac{1}{g v^2}\vec{\phi}_{vac}&\times \partial_M
\vec{\phi}_{vac} +\frac{1}{v}\vec{\phi}_{vac}A^0_M
\end{split}
\ee%
where the arrows denote the (adjoint) gauge indices and $A^0_M$ is
an arbitrary vector, and singlet of the gauge group. With the
above it is easy to check that $\partial_M \vec{\phi}_{vac}+i g
\vec{W}_M\times \vec{\phi}_{vac}=0$. The ``magnetic'' monopole
$U(1)$ gauge field strength is then given by $\bar F_{MN}=\frac{1}{v}
\vec{\phi}_{vac}\cdot F_{MN}$. This monopole solution is obviously
singular. We choose the three components of $\vec\phi$ to be given
by $v(\sin \theta \cos m\varphi, \sin \theta \sin m\varphi,
\cos\theta),\ m\in \mathbb{Z}$.

The stability analysis in this case has been carried out in
\cite{Nair} for the \emph{s}-wave perturbations (which corresponds to $m=1$
in this case). In fact the gauge transformation which brings $\phi$
to the direction of $z$-axis will map $W$ to our $\bar A$ of \eqref{monopole}
with $m=2n$.\footnote{It is known that with the adjoint Higgs one will only obtain Dirac monopoles of even charge.
In particular the lowest possible charge in this setting corresponds to $n=2$ in the notations of \eqref{monopole}.
This is the reason that our \emph{s}-wave is obtained for
$n=2$.} Repeating the analysis of the spectrum of gauge field
fluctuations about the above when Higgs is turned on, will therefore
produce the same result given by equation \eqref{e.o.m-No-grav.} but
the $M^2$ is now replaced
with%
\be\label{mass-opt-Higgs}%
 M^2 = - \{( D_m^2 -1)+2n\}+r^2g^2_{YM} v^2
  \ee%
where the last term is due to the Higgs giving mass to the gauge
fields. It can be checked that for small enough $r$, $r\lesssim
r_m$, with%
\be\label{monopole-size}%
r_m\sim (g_{YM}v)^{-1}%
\ee%
the last term becomes subdominant and we  again find the tachyonic
instability.

Finally we consider the {\it regular} 't Hooft-Polyakov solution.
The (approximate) form of solution may be found in eq.(29) of \cite{Luis-Hassan}.
This solution at large $r$ reduces to
\eqref{singular-monopole-Higgs}. At $r\lesssim r_m$, however, the
solution behaves differently and in particular at $r=0$ it is
regular. In this case the monopole is an object of effective radius
of $r_m$, rather than a point like object. The stability analysis for this case, due to the
$r$-dependence of the background Higgs and the gauge fields becomes very messy, however, one expects this solution
to be stable. This expectation is supported by topological reasons and, in our setting, by the
observation that  due to the monopole profile, the $M^2$ has a
non-trivial $r$ dependence and unlike the previous case, does not
change sign around $r_m$.

Similar arguments could be made for the fundamental Higgs case and
when we have dyons.

\subsection{The gravitational setting, instability in EYM(D) theoreis}\label{section2.2}

In this section we give a brief review of the non-Abelian
gravitating solutions of EYMD theory in four dimensions (for a more
detailed review see \cite{Volkov-Review} and references therein).
These solutions can be either point like, globally regular,
asymptotically flat (solitonic) or of the blackhole type. There are
a number of no-go theorems assuring that stationary solitonic
solutions do not exist in pure gravity or in YM theory. In gravity,
this is referred to as the Lichnerovicz's theorem which follows from
the fact that gravitational systems are purely attractive and hence
there is no way to make a balance of forces to keep the
gravitational soliton stationary. On the other hand the YM system is
purely repulsive and therefore it cannot support what is called a
``classical glueball''.

When combined together, gravity and YM can in principle make a
balance between forces to support stationary neutral
solitonic\footnote{There is a no-go theorem stating that there
exists no charged soliton in EYM \cite{Ershov}.} \cite{Bartnik} or
neutral and charged blackhole solutions \cite{SWY}. Similar class of
solutions have also been constructed for systems including a
dilatonic scalar field i.e. EYMD \cite{Lavrelash}.\footnote{Systems
including scalar fields in the non-trivial representation
(fundamental or adjoint) of the gauge group, the Higgs field, have
also been extensively studied under the name of EYMH \cite{ Nair, Lohiya,
Ortiz, Forgacs:2005nt}.}

In this section we will take the gauge group to be $SU(2)$ and
consider four dimensional, spherically symmetric, asymptotically
flat static solutions only. To summarize the results already
discussed in the literature and introduce the terminology used,  it
is convenient to start with the most general ansatz for {\it
spherically symmetric} solutions%
\be\label{GR-ansatz}%
\begin{split} ds^2&=\sigma^2N(dt+\alpha
dr)^2-\frac{1}{N}dr^2-R^2(d\theta^2+\sin^2\theta
d\phi^2)\\%
W&=aT_3+Im(wT_+)d\theta-\frac{n}{2} Re(wT_+)\sin\theta d\phi+
\frac{n}{2} T_3\cos\theta d\phi%
\end{split}\ee %
where $T_i\ (i=1\cdots 3)$ are the $SU(2)$ generators and
$T_+=T_1+iT_2$. Furthermore, $a$ is a real one form $a=a_0dt+a_rdr$
and the complex scalar $w$ as well as $\sigma, N, \alpha$ and $R$
only depend on $(t,r)$. The integer $n$ is the monopole charge.

 A study of the equations of motion for the YM field indicates that
for $n\ne2$ one should necessarily have $w=0$. Solutions of this
type are called {\it embedded Abelian }(eA), as the potential in
this case can be written in terms of the $U(1)$ potential ${\cal
A}= Tr(W T_3)=a+\frac{n}{2}\cos\theta d\phi$. The stationary (eA)
solutions are hence, always charged and are blackholes, the
so-called {\it colored blackholes}.

On the other hand, the {\it non-Abelian } (nA) solutions which can
only be achieved when $n=2$, may be solitonic or blackholes, but
always neutral. The stability of these (nA) solutions  against
{\it spherically symmetric perturbations} has been extensively
studied in the literature with the overall result that these
solutions are all unstable. There are two classes of the unstable,
tachyonic, modes; those in which the gravity fluctuations are
turned off, the {\it topological} or {\it odd parity} modes, and
those which also involve  metric perturbations, the {\it
gravitational} or {\it even parity} modes.

The above system can be generalized to the one which is more
relevant to string theory, by adding a dilaton field to the EYM to
end up with the EYMD system. The  dilaton, $\phi$, couples to the
YM action in the form of $e^{2\gamma\phi}L_{YM}$ where
$\gamma\ge0$ is a parameter. For any given $\gamma$, the (nA)
solutions of the EYMD system as well as their stability behavior
are the same as those of EYM, {\it i.e.} there are both odd and
even parity tachyonic modes \cite{Lavrelash, EYMD-SU(2)}. The
extreme limits of $\gamma\rightarrow0$ and
$\gamma\rightarrow\infty$ respectively correspond to the EYM and
YMD (in flat space). The latter which is essentially the case we
analyzed in section \ref{section2.1} has the same features of EYM
i.e. the dilaton, providing a purely attractive system, can
replace gravity leading to the same sort of solutions. The EYMD
will thus interpolate between these extreme systems and has the
same features of the two.
%Finally, the $\gamma=1$ case is the one that arises from
%certain compactifications of string theory.

Another system of interest is an EYM with a cosmological constant;
EYM-$\Lambda$ \cite{EYMDL}. The solutions of this system are
asymptotically (A)dS for (negative) positive values of $\Lambda$.
For $\Lambda>0$, the moduli space of (nA) solutions is essentially
that of the EYM system and all the solutions are still unstable,
with the same tachyonic modes. For $\Lambda<0$, however, the
situation is different. The main difference is that, the presence
of negative $\Lambda$ allows for a new continuous family of {\it
non-Abelian charged solitonic or blackhole} solutions. The
stability analysis for this case shows that \cite{negL,
Breitenlohner:2003qj}, although the solutions are still
generically unstable, we now have the possibility of having stable
solutions which of course do not have any counterpart in the
$\Lambda=0$ theory.

%perturbations. In the even parity sector, however, this analogy
%does not hold any longer. In fact it has been shown
%\cite{Breitenlohner:2003qj} that the moduli space of solutions
%consists of the following families: $(0,0)_0, (0,1)_0, (k,k)_k$
%and $(k,k-1)_k$ where $k=1,2,\cdots$ and $(a,b)_c$ refers to
%solutions with $c$ nodes which have $a(b)$ tachyonic modes in the
%odd(even) parity sector. The first family is thus stable against
%both kinds of perturbations.

%\fixme{Note that the GR people have only considered the "s-wave"
%fluctuations whereas discussions of previous section goes beyond
%the s-wave. In general, if we have a monopole of charge $n$, for
%even $n$ the fluctuations are tensors of $SU(2)$ rotation group
%and always there is an scalar mode, the ``s-wave''. If we have a
%monopole of odd charge, however, the fluctuations are spinorial
%and there is no ``s-wave''.}

\section{Instability of EYM(D) solutions with magnetic flux}\label{section3}

In the previous section we reviewed the instability of the so-called
non-Abelian solitonic or blackhole solutions. In this section we  apply the analysis of section 2.1 to
all possible {\it spherically symmetric} solutions to the EYM action
\be\label{EYM-action}%
 S=\int d^4x \sqrt{-g}(\frac{1}{8\pi G_N} R + Tr
 F_{MN}F^{MN})
%- g(\phi)_{ij}\nabla_M \phi^i\nabla^N\phi^j -\Lambda)
\ee%
when the magnetic flux ${\bar F}_{mn}$ \eqref{BG-F} is turned
on. With the gauge fields of the form of \eqref{BG-F} and noting
that $\bar F$ only takes values in the Cartan subalgebra of the
gauge group $G$, finding  solutions of the field equations is
essentially identical to the one in Einstein-Maxwell theory. These
solutions, if asymptotically flat, are of the form of (magnetized)
Reissner-Nordstr\"om (RN) blackholes of arbitrary magnetic charges
$n_i$. There is only one other class of solutions which are not
asymptotically flat. These are non-singular and of the form of
$AdS_2\times S^2$. The latter may be obtained as the near horizon
geometry of the extremal RN solutions.

We show that turning on gravity does not remove the instabilities we discussed in section \ref{section2.1} and
the same modes of the gauge field which were tachyonic in the non-gravitational case\footnote{In the RN solutions
generically there are four parameters (or quantum numbers) appearing, the Newton constant $G_N$, the Yang-Mills coupling
$g_{YM}$,
and monopole charges $n_i$ and the ADM mass of the blackhole. By non-gravitational limit we mean $G_N\to 0$ keeping the other
parameters fixed.} remain tachyonic.
Hence, we are led to the conclusion that
\vspace*{-.3cm}
\begin{center}{\it the presence of gravity is irrelevant to
the instability results of the non-gravitational case. Inclusion of
dilatonic scalar fields does not alter this result. Addition of Higgs scalars,  being a charged field under YM,
can however push some or all of the tachyonic modes inside the horizon and thus
may stabilize the system.}
\end{center}

In section \ref{section3.1} we analyze the (in)stability of embedded
Abelian, colored blackhole case. In comparison to the extensive
literature on the non-Abelian solutions, the literature on the
embedded Abelian case is slim and only the case with \emph{s}-wave
Gerogi-Glashow monopole has been considered in detail, showing that
such perturbations lead to instabilities
\cite{Nair}.\footnote{Reference  \cite{Nair} also considers some
tensorial non spherically symmetric modes. As discussed in section
\ref{section2.1}, however, for a generic monopole of charge $n$,
e.g. in the $SU(2)$ gauge theory, the spin of the tachyonic mode is
$n/2-1$. Our discussion applies to modes belonging to any
representation of the rotation $SU(2)$, tensorial as well as
spinorial.} In section \ref{section3.2} we study the instability of
the $AdS_2\times S^2$ solution. In section \ref{section3.3} we consider EYMD theory
and discuss the implication of the instability to the attractor mechanism.

\subsection{Instability of generic $4d$ colored blackholes}\label{section3.1}

The (in)stability analysis of  a colored, magnetized RN  blackhole
amounts to studying \eqref{e.o.m-No-grav.} with the proper choice for
the metric $g_{\mu\nu}$. In order that we take the metric of the form
 \be%
ds^2=-fdt^2+f^{-1}dr^2+b^2d\Omega_2^2%
\ee%
 where $f$ and $b$ are functions of $r$. Equation \eqref{e.o.m-No-grav.} written with this metric yields%
\be\label{e.o.m.with.grav.} %
  -f^{-1}\partial^2_t V_{+}+\partial_r(f \partial_r V_{+}) - \frac{1}{b^2} M^2V_{+} =0
  \ee%
  where $M^2$ has the same expression as in \eqref{mass-opt}. For
  concreteness, we take the gauge group to be $SU(2)$ for which
  the spectrum of $M^2$ has a simple form.
To analyze the above equation,  define the new coordinate $\rho$
\be%
d\rho = f^{-1}dr.%
\ee%
In this new coordinate system the metric becomes $
 ds^2 = f ( -dt^2 + d\rho^2).$
 Replacing
\[
V_{+}=e^{iEt}\psi(\rho),
\] %
 equation \eqref{e.o.m.with.grav.} then takes the form of a Schr\"odinger equation%
\be\label{Schrod.}%
  -\partial_\rho^2\psi  + U(\rho) \psi =E^2\psi,
 \ee%
 with the energy $E^2$ and the potential $U(\rho)$ %
\be\label{potential-f}%
U(\rho)= \frac{f}{b^2}M^2.%
\ee%

The question of instability, like the flat space case, therefore
reduces to the existence of normalizable, negative $E^2$ states (the
bound states).
The normalizability condition should now be imposed taking the non-flat
background metric into account. That is,
\be\label{Norm}%
\int d\rho \sqrt{f} b^2 |\psi|^2 < \infty.
\ee%
For the asymptotically flat case, like ours, at large $r$ (where the normalizability
concerns may arise) the measure reduces to that of standard flat space $r^2dr$.
This question could be answered once the function  $f$
is specified. (One should also note that functions $f$ and $b$
should be written in terms of $\rho$.) In the cases of interest for
us we are dealing with blackholes and hence the function $f(r)$
vanishes at the horizon and generically we have inner and outer
horizons, respectively $r_-$ and $r_h$. The Schrodinger equation
should be analyzed for $r\geq r_h$. The wave-function $\psi$ should
vanish for $r < r_h$, as the region inside the horizon is cut-off
from the rest of the space.

As the first example let us consider the {\it extremal}
Reissner-Nordstr\"om (RN) case, where%
\be\label{ERN-f.b.}%
f=(1- \frac{Q}{r})^2, \quad\quad\quad  b=r%
\ee%
$Q$ and the monopole charge $n$ are related as%
\be\label{charge}%
Q=\frac{n}{2g_{YM}} \sqrt{4\pi G_N}\equiv \frac{n}{2g_{YM}} M^{-1}_{Pl}.%
\ee%
In this case the explicit relation between $\rho$ and $r$ is%
\be\label{rho-r}%
 \rho= r -Q + 2Q ln\frac{r-Q}{Q} - \frac{Q^2}{r-Q}
\ee%
where,  for  convenience, we have chosen the integration constant
such that  $r= 2Q$ corresponds to $\rho =0$. Note that $\rho$
ranges from $-\infty$ to +$\infty$ as $r$ ranges from the horizon
$r_h=Q$ to $+\infty$. This coordinate does not cover the region
inside the horizon. The $(\rho,t)$ coordinate system is usually
called the Regge-Wheeler coordinates.

\begin{figure}[ht]
\begin{center}
\includegraphics[scale=0.8]{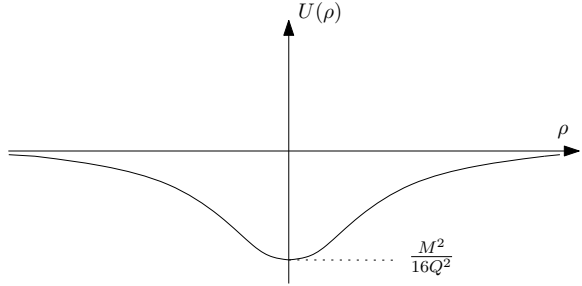}\caption{The Potential given in eq.\eqref{potential-f}
for the Extremal Reissner-Nordstr\"om blackhole, with $M^2=-n/2$.
This potential has negative energy bound states.} \label{Fig1}
\end{center}
\end{figure}

  It is easy to see that
$\rho= \pm \infty$ are inflection points of the potential and that
$U$ vanishes at these points. Furthermore, $U(\rho)$ has one more
extremum at $\rho=0$ which is a minimum for negative $M^2$ and a
maximum for positive $M^2$ and its value  at this extremum is
$\frac{M^2}{16Q^2}$. $U(\rho)$ is smooth everywhere and
$U(\rho)=U(-\rho)$. The case of interest for us is of course when
$M^2$ is negative which is depicted in Fig.\ref{Fig1}.

Qualitatively our quantum mechanical problem is very similar to the
motion of a particle in a potential of the form $\frac{M^2}{cosh^2
\rho}$ (the latter has been discussed in \cite{Landau}). This
potential has a finite number of of normalizable discrete modes
(bound states) for $ \frac{M^2}{16Q^2}\leq E^2\leq 0$.  For $E^2
\leq \frac{M^2}{16Q^2}$ of course there will be no solutions.
Existence of negative energy bound states, as discussed earlier,
means that the unstable modes, {\it i.e.} the modes which grow exponentially
in time,  which were also present in the flat background survive even
when the back reaction of the magnetic monopole on geometry is taken
into account.

One may repeat the above analysis for more general non-extremal
cases where
\[
f(r)=\frac{1}{r^2}(r-r_-)(r-r_h).
\] %
The qualitative behavior of the potential is essentially the same
as the extremal case. That is, for negative $M^2$ the potential
$U(\rho)$ has a minimum outside the horizon, and it is always
negative. Furthermore, in the $\rho\rightarrow \pm\infty$,
$U(\rho)$ asymptotes to zero.

\begin{figure}[ht]
\begin{center}
\includegraphics[scale=1]{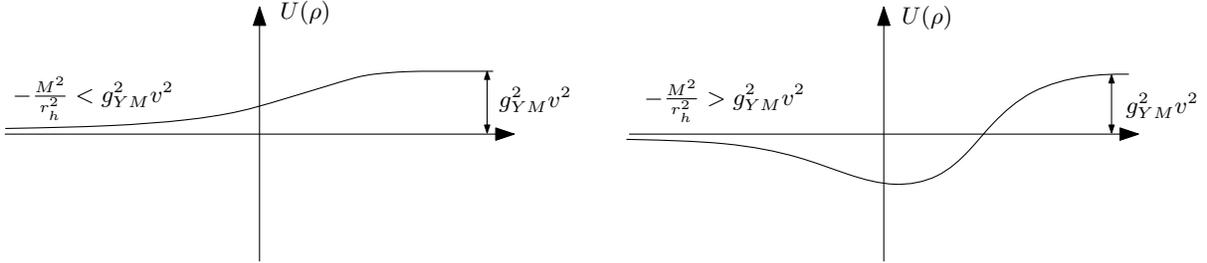}\caption{The potential for generic RN blackhole with
the Higgs. As we increase the Higgs expectation value $v$ the
minimum of the potential is pushed to the left and eventually can
be removed. The Figure on the left show the case with a large
Higgs and the right one shows the case with small $v$.}
\label{Fig3}
\end{center}
\end{figure}

Let us now add the non-singlet scalar field, the Higgs field. As discussed in section
\ref{section2.1}, this is done by replacing the ${M^2}$ by
${M^2}+g^2_{YM}{v^2}{b^2}$ ({\it cf.} eq.\eqref{mass-opt-Higgs}).
$U(\rho)$ becomes,%
\be\label{potential-Higgs}%
U(\rho)=\frac{(r-r_-)(r-r_h)}{r^2}({g^2_{YM}v^2}+\frac{M^2}{r^2})%
\ee%
where $v$ is the Higgs expectation value and $\rho$ and $r$ are
related as in \eqref{rho-r}. The potential $U(\rho)$, unlike the
previous case does not necessarily have bound states. As the
expectation value of the Higgs field increases, the minimum of
$U(\rho)$ is pushed behind the horizon. This can eventually result
in a potential without any bound states and hence large Higgs $vev$
can remove the instability\footnote{This observation had already
been made in \cite{Nair} for the \emph{s}-wave instabilities.}. Intuitively
and as a rough measure, when we add Higgs there are two length
scales in the problem, one is the horizon size $r_h$ and the other
is the monopole size $r_m$ \eqref{monopole-size}. When $r_h>r_m$ the
monopole is completely sitting behind the horizon and hence the
instability analysis is similarly to a ``point'' charge. In this
case, although the potential has a minimum, its minimum is sitting
behind the horizon and hence do not form a bound state. Therefore,
we do not see the instability. When $r_m>r_h$ we have a smoothed
out, extended monopole. As has been  depicted in Fig.\ref{Fig3} in
this case the potential does have the possibility of bound state,
and hence the instability. The same qualitative features remain when
we consider non-extremal case with the Higgs. The above potential
generalizes the results of \cite{Nair} to the case with arbitrary
monopole charge.

It turns out that bound states and
hence unstable modes appear if $M^2<0$ and if%
\be%
\frac{g_{YM}}{2\pi}M_{bh}<\frac{c\sqrt{-M^2}M_p^2}{v}+\frac{v}{c\sqrt{-M^2}}\frac{n^2}{4}%
\ee%
where $M_p$ is the planck mass and $M_{bh}=\frac{r_-+r_h}{2G}$ is
the ADM mass of the blackhole. $c$ is a positive number smaller
than one and arises from the details of quantum mechanics analysis
of having a bound state. For the extremal case, $M_{bh}G=Q$, $c$
approaches one.

\subsection{Instability in $AdS_2\times S^2$
background}\label{section3.2}

In this part we consider the other class of solutions to the EYM
theory in which the field strength is given by \eqref{BG-F}, the $AdS_2\times S^2$.
This solution can be obtained as the near horizon geometry of extremal colored blackholes and
hence this case is of particular interest because of its possible relevance to the
attractor mechanism \cite{Kallosh, Attractor-Triv2}.
It has been  discussed that the instability in such systems is
completely determined by the large  distance behavior of the
radial equation  \cite{Lohiya}. In this section we show that the
instability we have discussed in the previous sections is also
seen in the near horizon geometry.

It is straightforward to check that the $AdS_2\times S^2$:
\be\label{AdS-S}%
 ds^2=R_1^2(-f(r)dt^2+f^{-1} dr^2)+ R_2^2 d\Omega_2^2%
 \ee%
 where $R_1$ and $R_2$ are respectively
radii of the $AdS_2$ and $S^2$ parts, and $f(r)$ is
\be%
\begin{split}%
f(r)&=r^2\qquad\qquad\ \  {\rm for\ the\ Poincare'\ patch}\\
 f(r)&=1+r^2\qquad\quad {\rm in\ the\ global}\ AdS {\rm\ coordinates}
\end{split}%
\ee%
when $R_1=R_2=Q$ is a solution to our EYM theory. This solution can be obtained
in the near horizon limit of the four dimensional extremal Reissner-Nordstr\"om blackhole of previous section.
Although for our case,  the EYM theory and when four dimensional cosmological constant is zero, always
$R_1=R_2$, in our analysis we keep the $AdS$ and the sphere radii arbitrary.

Our instability analysis  goes through the same as before. We end up
with looking for the possible bound states of the Schr\"odinger
equation \eqref{Schrod.}, in which the potential is given by
\eqref{potential-f} with $b$ replaced by $R_2$. That is,
\footnote{Note that the $AdS_2$ case is special in the sense that it
has two disconnected one dimensional boundaries \cite{AdSfrag}. In
our case this means that the range of $\rho$ coordinate can be
extended to the full range of $(-\infty, +\infty)$ in the Poincare'
patch, and to $(-\frac{\pi}{2}, +\frac{\pi}{2})$ in the global
$AdS_2$. The boundaries are then located at $\rho=0$ for the former
and at $\rho=\pm\frac{\pi}{2}$ for the latter.}
\be\label{AdS-potential}%
U(\rho)=\left\{\begin{array}{ccc}%
&\frac{R_1^2}{R_2^2} M^2\ \frac{1}{\rho^2}\qquad
-\infty<\rho<+\infty\quad  \cr &     &     \cr &\frac{R_1^2}{R_2^2}
M^2\ \frac{1}{\cos^2\rho}\qquad -\frac{\pi}{2}<\rho< +\frac{\pi}{2}
\quad  &{\rm in\ the\ global}\ AdS {\rm\ coordinates}
\end{array}\right.%
\ee%

\begin{figure}[ht]
\begin{center}
\includegraphics[scale=0.9]{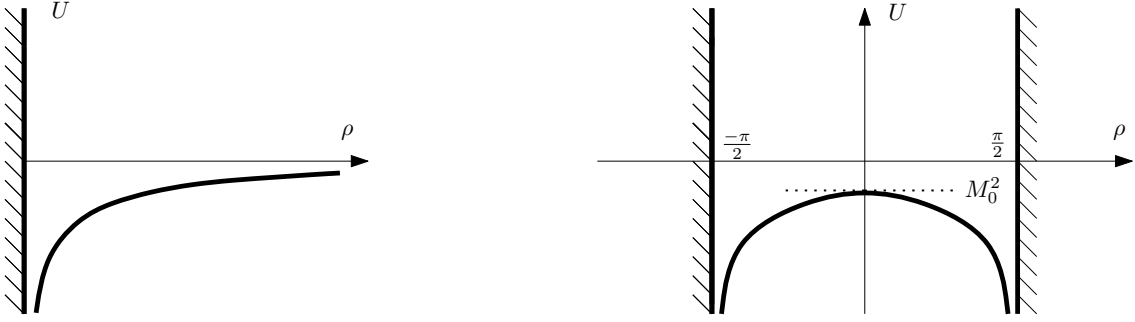}\caption{The potentials for
the $AdS_2$ case when $M_0^2=\frac{R_1^2}{R_2^2}M^2$ is negative.
The left figure corresponds to the Poincare' patch (as well as to
the flat space) and the one on the right corresponds to the global
coordinate. }\label{ads}
\end{center}
\end{figure}

To explore the bound states and discuss the instability we  need to
impose the correct normalizability and boundary conditions.  In the
case where we have a causal boundary, like the $AdS_2$ case, the
suitable boundary condition for the wave function is that of a
particle inside a box, with the walls of the box at the position of
the boundary, {\it i.e.} $\rho=0$ in the Poincare' patch and
$\rho=\pm\pi/2$ for the global $AdS_2$. The form of the potential
for negative $M^2$ is depicted in Fig.\ref{ads}. The normalization conditions read
\[%
\int_{0}^\infty d\rho\ \frac{1}{\rho} |\psi|^2<\infty\qquad  {\rm for\ the\
Poincare'\ patch}
\]%
and
\[%
 \int_{-\pi/2}^{\pi/2} d\rho\ \frac{1}{\cos\rho}\ |\psi|^2<\infty  \qquad  {\rm in\ the\ global}\ AdS {\rm\
 coordinates}.
\]%

The above Schr\"odinger equation is in fact exactly the same as the
wave equation for a massive scalar field of mass squared
$M_0^2=\frac{R_1^2}{R_2^2}M^2$ in the $AdS_2$.  Normalizable
negative energy states exist if $M^2_0<-1/4$. To see this, let us
for example consider the Schr\"odinger equation for the Poincare'
coordinates and focus on the equation around $\rho=0$. The solutions
can be of the form $\psi\sim \rho^\beta$, where $\beta$ satisfies
\eqref{e.o.m.with.grav.} with $M^2$ replaced by $M^2_{0}$. As in the
flat space case in order to have instability, $\beta$ should be a
complex number. The above discussion results in the well-known fact
that, to have tachyonic instability coming from a minimally coupled
scalar in an $AdS$ background, we need to {\it violate} the
Breitenlohner-Freedman bound \cite{BF}, which for the $AdS_2$ case
is exactly $M^2_0<-1/4$. For the case when the four dimensional
cosmological constant vanishes, $R_1=R_2$ and hence $M^2_0=M^2$. If
we choose our non-Abelian gauge group to be $SU(2)$, with a monopole
charge $n$ $(n>1)$, $M^2_0=-n/2$ ({\it cf}. discussions of section
2.1) and hence we have tachyonic instability.

Here we briefly address the case $R_1\neq R_2$. This happens when we
study near horizon limit of an extremal charged blackhole in a
background of non-vanishing four dimensional
cosmological constant $\Lambda$ \cite{Attractor, entropy-function}.
We then have
\[%
R_2^{-2}-R_1^{-2}=2\Lambda, \qquad Q^2={R_2^2}(1-R_2^2{\Lambda})
\]%
where $Q$ and the monopole charge $n$ are related as in
\eqref{charge}. Therefore, by tuning up $|\Lambda|$ we can increase
the $R_2/R_1$ ratio. When $\frac{R_1^2}{R_2^2}$ ratio exceeds $1/2n$
we do satisfy the Breitenlohner-Freedman bound and hence there is no
tachyon instability. More detailed and exhaustive analysis of this case, when a
non-zero cosmological constant is turned on, is postponed to future
works \cite{Progress}.

It is readily seen that for both cases we do have bound states and
hence the persistence of the instability. It is remarkable that in
the Poincare' patch the potential in terms of $\rho$ has exactly the
same expression as the similar case in the flat space ({\it cf.}
equation \eqref{e.o.m.with.grav.}). Note also that, it is $AdS$ in
the Poincare patch that is obtained as the direct result of taking
the near horizon limit. This may give a direct realization of the
observation we made in the beginning of this section.

As another remark we would like to note that in taking the near
horizon limit we are replacing the localized charges with fluxes
(see e.g. \cite{entropy-function}). This in particular implies that
the instability result is not limited to the charges and
``monopole'' configuration, but is a generic feature for the cases
we are dealing with non-vanishing non-Abelian  fluxes.

\subsection{Instability in EYMD and implications for the
attractors}\label{section3.3}

So far we have mainly focused on the EYM theory with a simple gauge
group $G$. We also discussed the EYMH theory and argued how addition
of a Higgs field  can change the (in)stability of the background
magnetic flux. Here we would like to address implications of
addition of the singlet scalar fields (the dilatons) to EYM. This is
the case of relevance to the attractor mechanism
\cite{Attractor-Triv}. The attractor mechanism simply states that
the near horizon geometry of the extremal charged blackholes only
depends  on the charges and is independent of the other details of
the theory, and in particular the moduli, if present
\cite{Attractor-Triv}. In other words, there are (infinitely) many
extremal blackhole solutions which reduce to the same solution in
the near horizon geometry and that
 the dynamics on the
horizon is decoupled from the dynamics of the rest of the space
\cite{Kallosh}. This result has been used to study the entropy of
the extremal blackholes \cite{entropy-function} and to identify
the micro-state counting of these blackholes \cite{DST}.

For the cases where we have a single dilaton field and a simple gauge group $G$,  presence of the dilaton field
does not essentially change the equations of motion for the gauge fields and following the
line of computations we have presented here, one can readily show that the same mode of the
gauge field fluctuation which we have shown to be tachyonic remains tachyonic in the
spherically symmetric magnetized solutions of  the EYMD theory.

In order to study the implications for the attractor mechanism,
recalling that when only the magnetic charges (fluxes) are turned on
one needs to have more than one $U(1)$ factors  (e.g. see
\cite{Attractor-Triv}), we consider a more general case of the EYMD
theory where our gauge group is a product of several simple factors
and also several dilaton fields are present: \be\label{EYMD-action}
 S=\int d^4x \sqrt{-g}(\frac{1}{8\pi G_N} R + Tr
F^a_{MN}F^{bMN}f_{ab}(\phi)
- g(\phi)_{ij}\partial_M \phi^i\partial^N\phi^j )%
\ee%
where $a,b$ run over $1,\cdots, k$ ($k$ is the number of simple
factors in the generic gauge group $G$),  $\phi^i$ are scalar
dilaton fields and the range of $i,j$ could be arbitrary,
independent of $k$, $g_{ij}$ is the metric on the moduli space of
the dilaton fields. Gauge invariance implies that $
f_{ab}(\phi)=e^{\alpha_{ai}\phi^i}\delta_{ab}. $

 The  extremal solutions of the above EYMD theory are
of relevance to the attractor mechanism \cite{Attractor-Triv}. For
concreteness let us choose the gauge group to be $SU(2)\times SU(2)$
({\it i.e. $k=2$}) and take a model with a single dilaton and
$f_{ab}(\phi) = \delta_{ab}e^{\alpha_a \phi}$, $a,b =1,2$.  The
exact attractor solution of reference \cite{Attractor-Triv} can be
readily embedded in this model. As shown in this reference for
$\alpha_1 \alpha_2 <0$ there exits an $\it exact$ attractor
solution. Assuming $\phi=const$, the monopole and blackhole
configuration of previous section will correspond to the zeroth
order solution of \cite{Attractor-Triv}. It is clear that the
analysis of sections 2.4.2 to 2.4.4 of this reference can be carried
through verbatim. The instability in non-Abelian flux backgrounds,
however, still remains there, when the background charge is
different from $n=1$ for each $SU(2)$. It is also straightforward,
along the line of section \ref{section3.2}, to check that the
$AdS_2\times S^2$ type solutions to the  above EYMD theory are also
unstable. In an obvious way one can show that the instability
remains for  (the extremal) solutions to the above EYMD
\eqref{EYMD-action} in its most general form.

As we discussed, the solutions to the EYMD theories which show the
attractor behavior, independently of the value of the dilaton fields
at infinity and the moduli, all suffer from the tachyonic
instability and the runaway behavior in the  small fluctuations of
some of the components of the gauge fields. Thus except for the
values of the magnetic charge $n=1$ for each $SU(2)$ factor, before
studying the attractor property, it is therefore necessary to
address the instability issue first.

%Therefore, as a result of the instability of the background, the
%very question of the attractor does not arise in the first place.%

\section{Relevance of colored blackholes to string theory}

We have discussed that EYMD theories  have unstable regular
or blackhole solutions. The instability manifests itself in the
modes with imaginary frequency. The EYM and YMD theories are two
specific limits of the EYMD theory. In this part we discuss the
relevance of EYMD theories to string theory.

\subsection{Non-Abelian gauge theories and Calabi-Yau compactifications}

Let us start with standard $10d$ superstrings and first consider the
heterotic strings compactified down to $4d$. The low energy
effective theory obtained in this way is generically a $4d$
(super)gravity theory plus some scalars (dilatons) and some gauge
fields. In the toroidal compactification, we start with $E_8\times
E_8$ or $SO(32)$ $10d$ heterotic strings and end up with ${\cal
N}=4$  supergravity in $4d$ with gauge group which is a subgroup of
$E_8\times E_8$ or $SO(32)$, plus 12 other $U(1)$ vector fields
coming from the ten dimensional metric and two form $B$-field upon
KK reduction \footnote{It is possible that in the self T-dual points
of the compactification moduli space this Abelian $U(1)^{12}$
enhances to non-Abelian $SU(2)^{12}$ or other larger groups of rank
$12$ \cite{Polchinski}.}. In a generic point of the moduli space,
where all the $16$ $E_8\times E_8$ or $SO(32)$ Wilson line moduli
are turned on, the gauge group is Higgsed down to $U(1)^{16}$
\cite{Polchinski}. Similarly, if we start with type I, type II or M
theories, upon toroidal compactification to $4d$, generally  we
obtain a $4d$ supergravity with $U(1)^{28}$ gauge group. Therefore,
at generic points of moduli spaces of string/M theory toroidal
compactifications non-Abelian gauge groups may be completely broken,
and hence the instability discussed here, does not show
up.\footnote{It is worth noting that closed strings on a torus, when
a background magnetic flux larger than some critical value is turned
on, contain tachyonic modes \cite{R-T}. It has been argued that this
tachyonic instability can signal a phase transition in the system
\cite{R-T}.} At non-generic points, where some of the Wilson line
moduli are not turned on, we do obtain non-Abelian gauge groups.

The toroidal compactification, although the simplest, is not the
most interesting string compactification. A phenomenologically
viable string compactification  should lead to non-Abelian GUTs or
the standard model. For this purpose the Calabi-Yau (CY)
compactification is a better choice. In this case, for the heterotic
or type I case, in the non-generic points of the compactification
moduli space, we obtain non-Abelian gauge groups which are subgroups
of $E_8\times E_8$ or $SO(32)$. In these cases we again face the
generic instability of the colored blackholes we discussed here.
Even if we are only interested in classical configurations which are
free of these instabilities, the fact that these states exist among
the classical solutions of the theory, at quantum level, can induce
instability.

\subsection{Non-Abelian gauge theories in G/H
compactifications}\label{section4.2}

The other class of compactifications which leads to non-Abelian
gauged supergravities in $4d$ are compactification on non-Ricci-flat
manifolds, e.g. $G/H$ coset manifolds,  $G$ being a compact group.
In these cases the isometries of the compactification manifold, $G$,
appears as the gauge group in the lower dimensional theory.
Moreover, generically we also obtain a non-vanishing cosmological
constant in the  lower dimensional theory. The value of the
cosmological constant is proportional to the scalar curvature of the
compactification manifold, though with the opposite sign. Hence, in
these cases we are dealing with (supersymmetric) EYMD-$\Lambda$
theory. As discussed in section \ref{section2.2}, appearance of
$\Lambda$  does not generically cure the instability of non-Abelian
solutions. There is, however, the possibility of having stable
non-Abelian solutions.  The EYMD-$\Lambda$ theory will be dealt with
in more detail in an upcoming publication \cite{Progress}.

As the last class  we mention the warped compactifications. Here the
non-Abelian gauge symmetry can appear in two ways. Either it arises
from the bulk gauge fields somehow confined to the branes
\cite{Parameswaran:2006db}, or from the internal degrees of freedom
on the branes, as in the Horava-Witten scenario \cite{HW} or its
generalizations and variants. In either cases we face the
instability discussed here and before using these setups for any
model building the instability issue should be addressed.

\section{Concluding remarks}

In this paper  we have studied the instability of general
spherically symmetric solutions of four dimensional  EYMD theories
with a constant (magnetic) field strength on the sphere. We showed
that the same mode which causes the instability in the gravitation
cases is also present in the non-gravitational case. The instability
is due to the fact that the lowest modes of gauge bosons in the
background of a Dirac magnetic monopole  on the flat space-time is
``tachyonic'' and has an exponential growth in time.

We showed that a Dirac magnetic monopole of a generic charge in a
non-Abelian gauge theory causes instability in the fluctuations of
the gauge fields. We should stress that, as discussed in section
\ref{section2.1}, although this statement is true for a generic
charge there exists cases corresponding to the minimum allowed value
of charges that there are no tachyons and are hence stable. For
example, for the $SU(2)$ gauge group, that is $n=1$. As discussed in
section \ref{section3}, these cases remain stable when gravity is
turned on.

One may wonder about the dynamics of the tachyon and ask for a
tachyon condensation mechanism with or without gravity.  One should
note that the tachyon is a feature of the linearized analysis. Its
exponential growth will eventually make the linearized analysis
invalid. A tachyon condensation of some kind may be the final fate
of this instability. However, for this an independent analysis is
called for.

An interesting outcome of our main result, which was indeed our
original motivation for looking into this problem, concerns the
attractor mechanism \cite{Attractor, Attractor-Triv} in extremal (supersymmetric or
non-supersymmetric) blackholes. These are charged blackhole
solutions to Einstein-Maxwell-Dilaton theory. According to the
attractor mechanism the near horizon behavior of the blackhole
solution is independent of the value of the dilatonic fields and
only depends on the charges defining the solution.  In the case of
the colored black holes, due to the instability discussed here, the
straightforward application of the attractor mechanism becomes
questionable.

The instability of the colored blackholes can also be relevant to
resolving the moduli problem in string compactification and removing
some of the pieces of the string landscape \cite{Lenny}. In a sense
this instability could be used as another  criterium for
distinguishing the swampland from the landscape \cite{swampland}.

Here we mainly focused on the EYM theory in four dimensions. One of
the cases which is of outmost interest is the EYM-$\Lambda$
theories. As briefly discussed in section \ref{section4.2} these
theories are very relevant to string theory. It is interesting to
study the (in)stability of the embedded Abelian, colored blackhole
solutions in the context of $4d$ gauged supergravities
\cite{Progress}. Among this class the case of $4d$ $SO(8)$ gauged
supergravity, which is related to M-theory on $S^7$ and the
stability of the $4d$ AdS-RN blackholes are of particular interest.
It will also be very interesting to generalize our analysis of
section \ref{section3} to the cases in higher dimensions and check
whether the tachyonic modes still persist. We would expect that the
kind of instability present in four dimensions should also appear in
higher dimensions. This expectation should, however, be checked by
explicit computations. One of the most interesting cases concerns
the charged blackhole configurations of the $5d$ $SU(4)$ gauged
supergravity, arising from $IIB/S^5$. These are the cases relevant
to the AdS/CFT duality \cite{Progress}.

\section*{Acknowledgements}

We are grateful to Gary Gibbons for bringing  some old references on
the subject of colored black holes and their instabilities to our
attention. We would like to express our gratitude to the organizers
of the Third Crete Regional Conference where this project was
initiated. A.E.M would also like to thank the high energy section of
the Abdus-Salam ICTP where some of this research was carried out.

%\vspace{3cm}


\begin{thebibliography}{99}
%\cite{Goldstein:2005hq}




\bibitem{Sen-94}
  A.~Sen,
  ``Black hole solutions in heterotic string theory on a torus,''
  Nucl.\ Phys.\ B {\bf 440}, 421 (1995)
  [arXiv:hep-th/9411187].
  %%CITATION = HEP-TH 9411187;%%

\bibitem{Bartnik}
  R.~Bartnik and J.~Mckinnon,
  ``Particle-Like Solutions Of The Einstein Yang-Mills Equations,''
  Phys.\ Rev.\ Lett.\  {\bf 61}, 141 (1988).
  %%CITATION = PRLTA,61,141;%%

  J.~A.~Smoller and A.~G.~Wasserman,
  ``Existence of infinitely many smooth, static, global solutions of the
  Einstein / Yang-Mills equations,''
  Commun.\ Math.\ Phys.\  {\bf 151}, 303 (1993).
  %%CITATION = CMPHA,151,303;%%

\bibitem{Perry75-7}
  P.~Van Nieuwenhuizen, D.~Wilkinson and M.~J.~Perry,
  ``On A Regular Solution Of 'T Hooft's Magnetic Monopole Model In Curved
  Space,''
  Phys.\ Rev.\ D {\bf 13}, 778 (1976).
  %%CITATION = PHRVA,D13,778;%%
M.~J.~Perry,
  ``Black Holes Are Colored,''
  Phys.\ Lett.\ B {\bf 71} (1977) 234.
  %%CITATION = PHLTA,B71,234;%%


\bibitem{SWY}
 J.~A.~Smoller, A.~G.~Wasserman and S.~T.~Yau,
  ``Existence of black hole solutions for the Einstein / Yang-Mills
  equations,''
  Commun.\ Math.\ Phys.\  {\bf 154}, 377 (1993).
  %%CITATION = CMPHA,154,377;%%

 P.~Breitenlohner, P.~Forgacs and D.~Maison,
``On Static spherically symmetric solutions of the Einstein Yang-Mills
 equations,''
  Commun.\ Math.\ Phys.\  {\bf 163}, 141 (1994).
  %%CITATION = CMPHA,163,141;%%




\bibitem{Lavrelash}
  G.~V.~Lavrelashvili and D.~Maison,
 ``Regular and black hole solutions of Einstein Yang-Mills Dilaton theory,''
  Nucl.\ Phys.\ B {\bf 410}, 407 (1993).
  %%CITATION = NUPHA,B410,407;%%



\bibitem{Volkov-Review}
  M.~S.~Volkov and D.~V.~Gal'tsov,
  ``Gravitating non-Abelian solitons and black holes with Yang-Mills  fields,''
  Phys.\ Rept.\  {\bf 319}, 1 (1999)
  [arXiv:hep-th/9810070].
  %%CITATION = HEP-TH 9810070;%%


\bibitem{RSS2}
  S.~Randjbar-Daemi, A.~Salam and J.~A.~Strathdee,
  ``Instability Of Higher Dimensional Yang-Mills Systems,''
  Phys.\ Lett.\ B {\bf 124}, 345 (1983)
  [Erratum-ibid.\ B {\bf 144}, 455 (1984)].
  %%CITATION = PHLTA,B124,345;%%

  %\cite{Dvali:2001qr}
\bibitem{DRT}
  G.~R.~Dvali, S.~Randjbar-Daemi and R.~Tabbash,
  ``The origin of spontaneous symmetry breaking in theories with large  extra
  dimensions,''
  Phys.\ Rev.\ D {\bf 65}, 064021 (2002)
  [arXiv:hep-ph/0102307].
  %%CITATION = HEP-PH 0102307;%%



\bibitem{Attractor}
  S.~Ferrara and R.~Kallosh,
  ``Supersymmetry and Attractors,''
  Phys.\ Rev.\ D {\bf 54}, 1514 (1996)
  [arXiv:hep-th/9602136].
  %%CITATION = HEP-TH 9602136;%%

\bibitem{Attractor-Triv}
  K.~Goldstein, N.~Iizuka, R.~P.~Jena and S.~P.~Trivedi,
  ``Non-supersymmetric attractors,''
Phys.\ Rev.\ D {\bf 72}, 124021 (2005),
 arXiv:hep-th/0507096.
  %%CITATION = HEP-TH 0507096;%%


\bibitem{Luis-Hassan}
  L.~Alvarez-Gaume and S.~F.~Hassan,
   ``Introduction to S-duality in N = 2 supersymmetric gauge theories: A
pedagogical review of the work of Seiberg and Witten,''
  Fortsch.\ Phys.\  {\bf 45}, 159 (1997)
  [arXiv:hep-th/9701069].
  %%CITATION = HEP-TH 9701069;%%

\bibitem{Nair}
 K.~M.~Lee, V.~P.~Nair and E.~J.~Weinberg,
%``Black holes in magnetic monopoles,''
%  Phys.\ Rev.\ D {\bf 45}, 2751 (1992)
%  [arXiv:hep-th/9112008];
  %%CITATION = HEP-TH 9112008;%%
``A Classical instability of Reissner-Nordstr\"om solutions and the fate of
  magnetically charged black holes,''
  Phys.\ Rev.\ Lett.\  {\bf 68}, 1100 (1992)
  [arXiv:hep-th/9111045].
  %%CITATION = HEP-TH 9111045;%%

\bibitem{Ershov}
  A.~A.~Ershov and D.~V.~Galtsov,
   ``Nonexistence Of Regular Monopoles And Dyons In The SU(2) Einstein
 Yang-Mills Theory,''
  Phys.\ Lett.\ A {\bf 150}, 159 (1990).
  %%CITATION = PHLTA,A150,159;%%

\bibitem{RSS1}
  S.~Randjbar-Daemi, A.~Salam and J.~A.~Strathdee,
  ``Spontaneous Compactification In Six-Dimensional Einstein-Maxwell Theory,''
  Nucl.\ Phys.\ B {\bf 214}, 491 (1983).
  %%CITATION = NUPHA,B214,491;%%


\bibitem{Lohiya}
  D.~Lohiya,
``Stability Of Einstein Yang-Mills Monopoles And Dyons,''
  Annals Phys.\  {\bf 141}, 104 (1982).
  %%CITATION = APNYA,141,104;%%

\bibitem{Ortiz}
  M.~E.~Ortiz,
  ``Curved space magnetic monopoles,''
  Phys.\ Rev.\ D {\bf 45}, 2586 (1992).
  %%CITATION = PHRVA,D45,2586;%%

P.~Breitenlohner, P.~Forgacs and D.~Maison,``Gravitating monopole solutions 1\& 2,''
  Nucl.\ Phys.\ B {\bf 383}, 357 (1992);
  %%CITATION = NUPHA,B383,357;%%;
%``Gravitating monopole solutions. 2,''
    Nucl.\ Phys.\ B {\bf 442}, 126 (1995)
    [arXiv:gr-qc/9412039].
    %%CITATION = GR-QC 9412039;%%


  \bibitem{Forgacs:2005nt}
 P.~Forgacs and S.~Reuillon,
  ``Spatially compact solutions and stabilization in  Einstein-Yang-Mills-Higgs
  theories,''
  Phys.\ Rev.\ Lett.\  {\bf 95}, 061101 (2005)
  [arXiv:gr-qc/0505007].
  %%CITATION = GR-QC 0505007;%%


\bibitem{EYMD-SU(2)}
  W.~H.~Aschbacher,
  ``On the instabilities of the static, spherically symmetric SU(2)
  Einstein-Yang-Mills-dilaton solitons and black holes,''
  Phys.\ Rev.\ D {\bf 73}, 024014 (2006)
  [arXiv:gr-qc/0509060].
  %%CITATION = GR-QC 0509060;%%

\bibitem{EYMDL}
  T.~Torii, K.~i.~Maeda and T.~Tachizawa,
  ``Cosmic colored black holes,''
  Phys.\ Rev.\ D {\bf 52}, 4272 (1995)
  [arXiv:gr-qc/9506018].
  %%CITATION = GR-QC 9506018;%%

 M.~S.~Volkov, N.~Straumann, G.~V.~Lavrelashvili, M.~Heusler and O.~Brodbeck,
  ``Cosmological Analogues of the Bartnik--McKinnon Solutions,''
  Phys.\ Rev.\ D {\bf 54}, 7243 (1996)
  [arXiv:hep-th/9605089].
  %%CITATION = HEP-TH 9605089;%%

\bibitem{negL}
  J.~Bjoraker and Y.~Hosotani,
  ``Stable monopole and dyon solutions in the Einstein-Yang-Mills theory in
  asymptotically anti-de Sitter space,''
  Phys.\ Rev.\ Lett.\  {\bf 84}, 1853 (2000)
  [arXiv:gr-qc/9906091];
  %%CITATION = GR-QC 9906091;%%
   ``Monopoles, dyons and black holes in the four-dimensional
  Einstein-Yang-Mills theory,''
  Phys.\ Rev.\ D {\bf 62}, 043513 (2000)
  [arXiv:hep-th/0002098].
  %%CITATION = HEP-TH 0002098;%%

  E.~Winstanley,
  ``Existence of stable hairy black holes in SU(2) Einstein-Yang-Mills  theory
  with a negative cosmological constant,''
  Class.\ Quant.\ Grav.\  {\bf 16}, 1963 (1999)
  [arXiv:gr-qc/9812064].
  %%CITATION = GR-QC 9812064;%%



\bibitem{Breitenlohner:2003qj}
  P.~Breitenlohner, D.~Maison and G.~Lavrelashvili,
  ``Non-Abelian gravitating solitons with negative cosmological constant,''
  Class.\ Quant.\ Grav.\  {\bf 21}, 1667 (2004)
  [arXiv:gr-qc/0307029].
  %%CITATION = GR-QC 0307029;%%

 P.~Breitenlohner, P.~Forgacs and D.~Maison,
``Classification of static, spherically symmetric solutions of the
  Einstein-Yang-Mills theory with positive cosmological constant,''
  Commun.\ Math.\ Phys.\  {\bf 261}, 569 (2006)
  [arXiv:gr-qc/0412067].
  %%CITATION = GR-QC 0412067;%%

P.~Forgacs and S.~Reuillon,
  ``On the number of instabilities of cosmological solutions in an
  Einstein-Yang-Mills system,''
  Phys.\ Lett.\ B {\bf 568}, 291 (2003)
  [arXiv:hep-th/0306126].
  %%CITATION = HEP-TH 0306126;%%


\bibitem{Landau}
Landau Text book Series, Quantum Mechanics, page 73, problem 5.

\bibitem{Kallosh}
  R.~Kallosh, N.~Sivanandam and M.~Soroush,
  ``The non-BPS black hole attractor equation,''
  JHEP {\bf 0603}, 060 (2006)
  [arXiv:hep-th/0602005].
  %%CITATION = HEP-TH 0602005;%%
\bibitem{Attractor-Triv2}
 K.~Goldstein, R.~P.~Jena, G.~Mandal and S.~P.~Trivedi,
  ``A C-function for non-supersymmetric attractors,''
  JHEP {\bf 0602}, 053 (2006)
  [arXiv:hep-th/0512138].
  %%CITATION = HEP-TH 0512138;%%

P.~Kaura and A.~Misra,
  ``On the existence of non-supersymmetric black hole attractors for
  two-parameter Calabi-Yau's and attractor equations,''
  arXiv:hep-th/0607132.
  %%CITATION = HEP-TH 0607132;%%







\bibitem{AdSfrag}
  J.~M.~Maldacena, J.~Michelson and A.~Strominger,
  ``Anti-de Sitter fragmentation,''
  JHEP {\bf 9902}, 011 (1999)
  [arXiv:hep-th/9812073].
  %%CITATION = HEP-TH 9812073;%%











\bibitem{BF}
  P.~Breitenlohner and D.~Z.~Freedman,
  ``Positive Energy In Anti-De Sitter Backgrounds And Gauged Extended
  Supergravity,''
  Phys.\ Lett.\ B {\bf 115}, 197 (1982).
  %%CITATION = PHLTA,B115,197;%%

 E.~Witten,
 ``Anti-de Sitter space and holography,''
  Adv.\ Theor.\ Math.\ Phys.\  {\bf 2}, 253 (1998)
  [arXiv:hep-th/9802150].
  %%CITATION = HEP-TH 9802150;%%

\bibitem{Progress}
A. E. Mosaffa, S. Randjbar-Daemi, M.M. Sheikh-Jabbari, ``{\it
%On the Irrelevance of Gravity to the Instabilities of EYM-$\Lambda$ Theories and its implications for
%four and five dimensional gauged supergravities}
%}
Instabilities in Magnetized Four and Five Dimensional Gauged Supergravity Solutions}
'', Work in Progress.


\bibitem{entropy-function}
  A.~Sen,
  ``Black hole entropy function and the attractor mechanism in higher
  derivative gravity,''
  JHEP {\bf 0509}, 038 (2005)
  [arXiv:hep-th/0506177].
  %%CITATION = HEP-TH 0506177;%%

\bibitem{DST}
  A.~Dabholkar, A.~Sen and S.~Trivedi,
``Black Hole Microstates and Attractor Without Supersymmetry,''
  arXiv:hep-th/0611143.
  %%CITATION = HEP-TH 0611143;%%

D.~Astefanesei, K.~Goldstein and S.~Mahapatra,
 ``Moduli and (un)attractor black hole thermodynamics,''
  arXiv:hep-th/0611140.
  %%CITATION = HEP-TH 0611140;%%

\bibitem{Polchinski}
 J. Polchinski, ``String Theory'', Vol.1 \& 2, Cambridge University
 Press, 1997.
\bibitem{R-T}
  J.~G.~Russo and A.~A.~Tseytlin,
  ``Constant magnetic field in closed string theory: An Exactly solvable
  model,''
  Nucl.\ Phys.\ B {\bf 448}, 293 (1995)
  [arXiv:hep-th/9411099].
  %%CITATION = HEP-TH 9411099;%%


\bibitem{Parameswaran:2006db}
  S. Parameswaran, S. Randjbar-Daemi and A. Salvio
  `` Guage fields, fermions and mass gaps in 6D brane words,''
  arXiv:hep-th/0608074.
  %%CITATION = HEP-TH 0608074;%%

\bibitem{HW}
 P.~Horava and E.~Witten,
  ``Heterotic and type I string dynamics from eleven dimensions,''
  Nucl.\ Phys.\ B {\bf 460}, 506 (1996)
  [arXiv:hep-th/9510209];
  %%CITATION = HEP-TH 9510209;%%
 %P.~Horava and E.~Witten,
  ``Eleven-Dimensional Supergravity on a Manifold with Boundary,''
  Nucl.\ Phys.\ B {\bf 475}, 94 (1996)
  [arXiv:hep-th/9603142].
  %%CITATION = HEP-TH 9603142;%%


\bibitem{Lenny}
  L.~Susskind,
  ``The anthropic landscape of string theory,''
  arXiv:hep-th/0302219.
  %%CITATION = HEP-TH 0302219;%%

\bibitem{swampland}
  C.~Vafa,
``The string landscape and the swampland,''
  arXiv:hep-th/0509212.
  %%CITATION = HEP-TH 0509212;%%
 H.~Ooguri and C.~Vafa,
  ``On the geometry of the string landscape and the swampland,''
  arXiv:hep-th/0605264.
  %%CITATION = HEP-TH 0605264;%%



\end{thebibliography}
\end{document}